\documentclass[twocolumn,showpacs,preprintnumbers,amsmath,amssymb]{revtex4} 
                                                                                
\usepackage{graphicx}
\usepackage{dcolumn}
\usepackage{bm}
\begin{document}
                                                                                
\preprint{CUPhys/13/2006}
                                                                                
\title{Floating Phase in 1D Transverse ANNNI Model}
                                                                                
\author{Anjan Kumar Chandra }
\author{Subinay Dasgupta}%
\affiliation{%
Department of Physics, University of Calcutta,92 Acharya Prafulla Chandra Road, Calcutta 700009, India.\\
}%
                                                                                
\date{\today}
                                                                                
\begin{abstract}
To study the ground state of ANNNI chain under transverse field as a function
of frustration parameter $\kappa$ and field strength $\Gamma$, we present
here two different perturbative analyses. In one, we consider the (known)
ground state at $\kappa=0.5$ and $\Gamma=0$ as the unperturbed
state and treat an increase of the field from 0 to $\Gamma$ coupled with
an increase of $\kappa$ from 0.5 to $0.5+r\Gamma$ as perturbation. The first
order perturbation correction to eigenvalue can be calculated exactly and we
could conclude that there are only two phase transition lines emanating from
the point $\kappa=0.5$, $\Gamma=0$. In the second perturbation scheme, we
consider the number of domains of length 1 as the perturbation and obtain
the zero-th order eigenfunction for the perturbed ground state.
From the longitudinal spin-spin correlation, we
conclude that floating phase exists for small values of transverse field over
the entire region intermediate between the ferromagnetic phase and antiphase.
                                                                                
\end{abstract}
                                                                                
\pacs{64.60.Fr, 75.10.Jm, 05.50.+q}
                                                                                
\def\be{\begin{equation}}
\def\ee{\end{equation}}
\maketitle

\section{Introduction}

\begin{figure}
\noindent \includegraphics[width= 8cm, angle = 0]{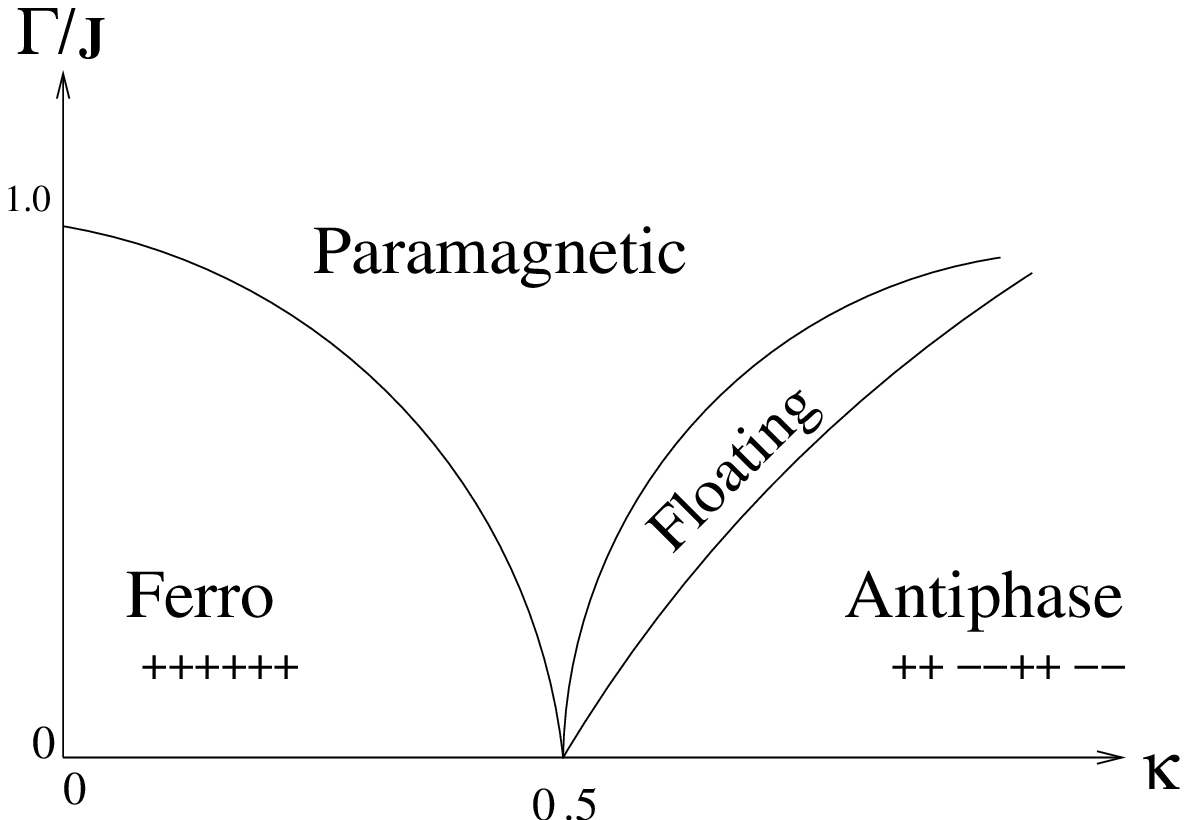}
\caption\protect{\label{fig:Phase} Schematic phase diagram of the transverse 
ANNNI model according to early investigations [1,2]. }
\end{figure}

The transverse  Axial Next-Nearest Neighbour Ising (ANNNI) model is one of 
the simplest Ising models that contains 
tunable frustration and tunable quantum fluctuation. In one dimension, it
is defined (for spin=$\frac{1}{2}$) by the Hamiltonian \cite{bkc_book}
\be
{\mathcal H} =  - J \sum_{j=1}^N (s^z_j s^z_{j+1} - \kappa s^z_j s^z_{j+2}) 
- \Gamma \sum_{j=1}^N s^x_j 
\ee
Here the first term describes a ferromagnetic nearest-neighbour interaction
of strength $J$($>0$) between longitudinal components of spin $s^z$ 
($=\pm 1$), the second term describes an antiferromagnetic second
neighbour interaction of strength $\kappa J$ also in the longitudinal 
direction, and the third term describes an external field in the transverse 
direction
of strength $\Gamma$. The ratio $\kappa$ ($>0$) is called the 
frustration parameter. This Hamiltonian describes a classical ANNNI 
chain subjected to a transverse field as well as a transverse 
(nearest-neighbour) Ising model 
with an additional frustrated second nearest-neighbour interaction. 
In this paper, we shall deal with the {\em ground state} (zero-temperature)
phase diagram of the Hamiltonian ${\mathcal H}$. 
It is known \cite{bkc_book,mattis} that for 
$\kappa=0$ (only nearest-neighbour interaction), there is a ferromagnetic to 
paramagnetic second order phase transition at $\Gamma=J$. On the other hand,
for $\Gamma=0$ (classical 
ANNNI chain) \cite{bkc_book,selke} the ground state is ferromagnetic for 
$\kappa < 0.5$ and antiphase
($++--$ type) for $\kappa > 0.5$, with a `multiphase point' at $\kappa = 0.5$.
At the multiphase point \cite{selke} the ground-state has very high 
($\sim g^N$, where $g$ is the golden ratio $(\sqrt{5}+1)/2$) degeneracy as any
combination of antiphase and ferromagnetic patches will serve as ground state
configuration. From approximate analytic and numerical approaches 
\cite{bkc_book,rieger}, early studies had proposed a phase diagram 
(Fig.~\ref{fig:Phase}) that consists of ferromagnetic, paramagnetic and 
antiphase regions,
alongwith a region of what is called {\em floating phase}. The $n^{th}$ 
neighbour spin-spin correlation function in the longitudinal direction
\be C^z(n) \equiv  < s^z_j s^z_{j+n} > - <s^z_j>^2 \ee
decays exponentially with distance in the ferromagnetic, antiphase and paramagnetic 
regions but decays algebraically in the floating phase. The evidence for
the presence of floating phase is provided by quantum 
Monte Carlo 
simulation \cite{ariz} and by exact numerical diagonalisation of small systems
\cite{sen,rieger}, one of which \cite{rieger} extends even upto a length of 32.

Recently, a question has arisen on the existence of the floating phase,
as a recent study \cite{amit} has claimed that the floating phase exists 
only within a strip of infinitesimally small width. 
The transverse ANNNI chain  is related to the 
two-dimensional classical ANNNI model by Suzuki-Trotter transformation. The
phase diagram for this model is also similar to Fig.~\ref{fig:Phase} (with $\Gamma$ replaced
by temperature). Here also the 
earlier studies \cite{selke} support the existence of floating 
phase in view of Monte Carlo simulations by Selke and others \cite{selke2} 
and approximate analytic calculations by Villain and Bak \cite{VB}, while a 
recent measurement of dynamical exponent \cite{jap} and
a density matrix renormalisation group analysis \cite{jap2}
claim that the floating phase exists, if at all, along a line only.
We have also studied \cite{cd1} some static and dynamical properties of the 
2D ANNNI model by Monte Carlo simulation. Our study also indicates that
the floating phase exists, if at all, only along a line.
The question is to whether the floating phase exists only along a line or 
extends over a region in the case of transverse ANNNI chain is the basic 
motivation of this paper. 

\begin{figure}
\noindent \includegraphics[width= 8cm, angle = 0]{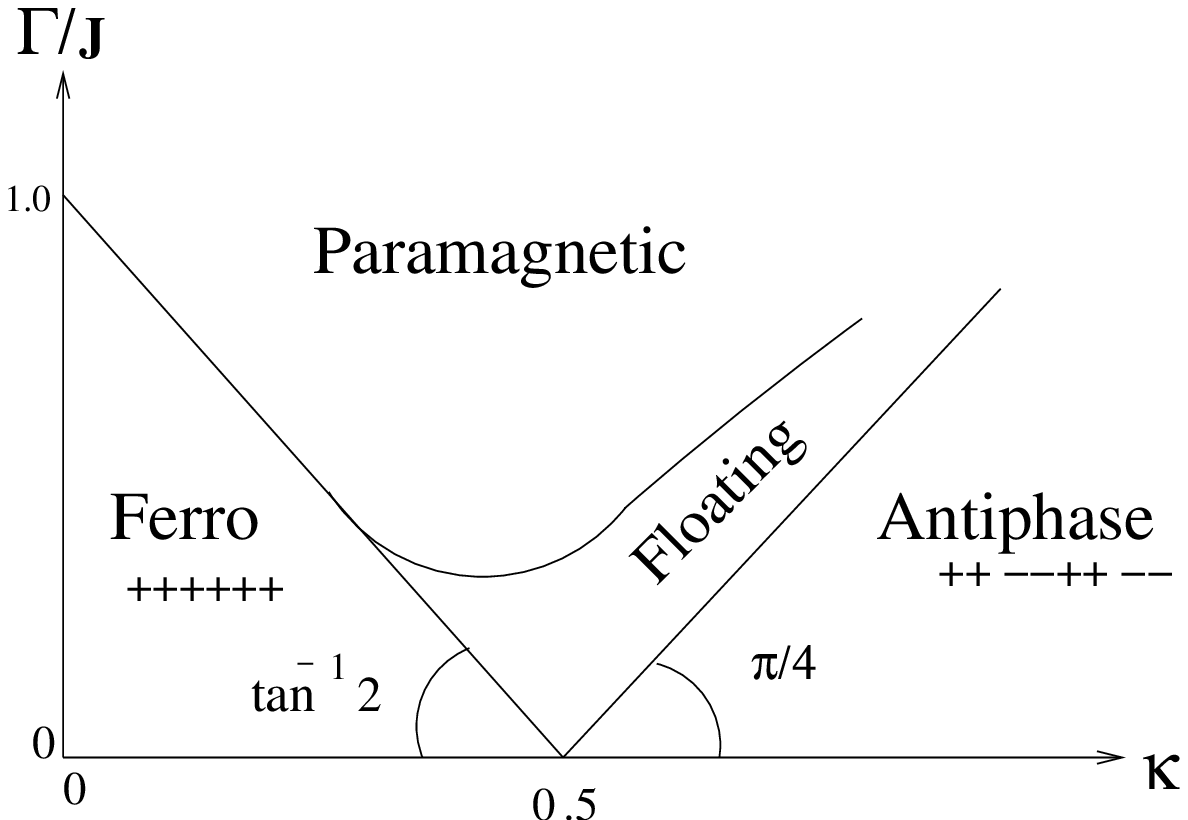}
\caption\protect{\label{fig:Phase2}Schematic phase diagram of the transverse ANNNI model
according to the present work. }
\end{figure}

In this paper we perform two perturbation calculations with an aim to have
an idea about the phase that exists near the multiphase point for non-zero 
transverse field. Our conclusion is that there exists floating phase over a 
region extending from ferromagnetic phase to antiphase (for 
small values of $\Gamma$) and the phase diagram looks like Fig.~\ref{fig:Phase2}.
This result is in contradiction with previous results as none of the previous
studies had predicted  floating phase for $\kappa < 0.5$. We must mention 
that all our results are true only at ``small'' values of $\Gamma$ and the 
quantitative details of this diagram is not reliable at $\Gamma \sim 1$. 
However, the topological structure of the diagram should be correct.
This article is organised as follows. In Sections II and III, the first and
the second perturbation schemes will be presented (respectively) and we shall 
conclude with some discussions in Sec. IV.

\begin{figure}
\noindent \includegraphics[width= 8cm, angle = 0]{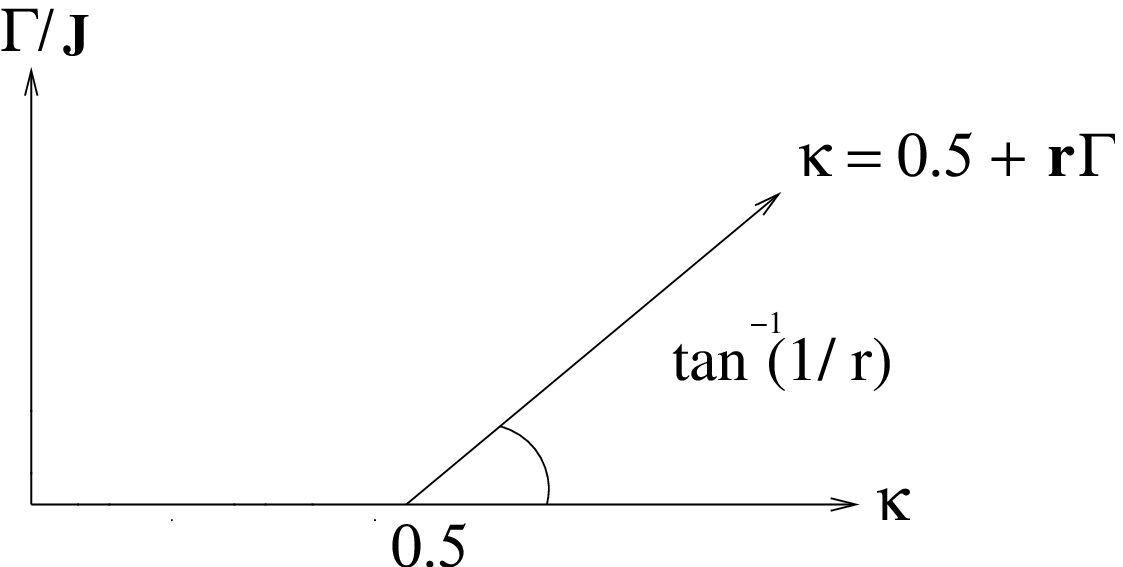}
\caption\protect{\label{fig:trans}Scheme of the first perturbation treatment. The transverse
field is
altered from 0 to a (small) value $\Gamma$ and the frustration parameter
$\kappa$ is altered from 0.5 to $0.5 + r\Gamma$. The parameter $r$ may vary
from $-\infty$ to $\infty$. }
\end{figure}
                                                                                
\begin{figure}
\noindent \includegraphics[width= 8cm, angle = 0]{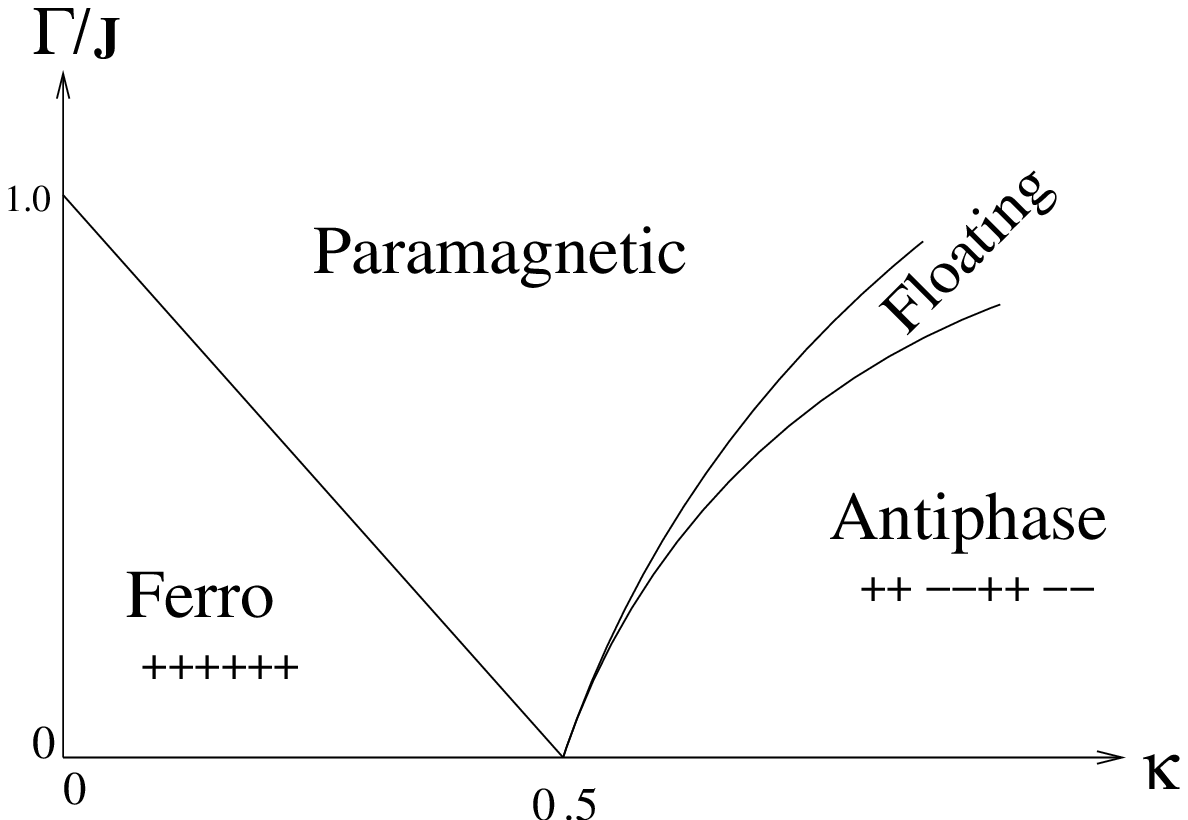}
\caption\protect{\label{fig:Phase3}One possible type of phase diagram of the 
transverse ANNNI model. }
\end{figure}

\section{The First Perturbation Scheme}

\subsection {Principle}

We shall now present a first order perturbation 
calculation around $\kappa=0.5$ by rewriting the Hamiltonian
$ {\mathcal H}$ of Eq. (1) as
\[
{\mathcal H} = {\mathcal H}_{cl} + {\mathcal H}_p
\]
where
\be
{\mathcal H}_{cl} = - J\sum_j [ s^z_j s^z_{j+1} - \frac{1}{2} s^z_j s^z_{j+2}]
\ee
and
\be
{\mathcal H}_p = \sum_j \Gamma \left[ - s^x_j + r s^z_j s^z_{j+2} \right]
\ee
with $r = J(\kappa -0.5)/\Gamma$. We always take $\Gamma$ to be positive, so
that negative values of $r$ imply $\kappa < 0.5$.

We shall treat the ${\mathcal H}_p$ as a perturbation over ${\mathcal H}_{cl}$.
Obviously, ${\mathcal H}_{cl}$ is the longitudinal (classical) part of 
${\mathcal H}$ at $\kappa=0.5$ and ${\mathcal H}_p$ includes the transverse 
part and an increase of $\kappa$ (from 0.5) by an amount $r\Gamma$. One must 
note that for a perturbative treatment $\Gamma$ has to be small
but $r$ need not be so. In the $(\kappa,\Gamma)$ phase space, the perturbation
takes us from the multiphase point (0.5, 0) to a point 
(0.5 + $r\Gamma$, $\Gamma$). As $r$ varies from $-\infty$ to $\infty$, the 
perturbation takes us from the ferromagnetic to the antiphase region (Fig.~\ref{fig:trans}).
The first order correction ($E^{(1)}$) to ground state energy can be
calculated {\em exactly}. This quantity is studied as a function of $r$, 
because if there is a non-analyticity at $r=r_0$, then there will be a phase 
transition line emanating from the (0.5, 0) point at an angle $\tan^{-1}(1/r_0)$
with the increasing $\kappa$ direction.
It is found that $E^{(1)}$ as a function of $r$ is 
non-analytic only at $r=-0.5$ and 1. Therefore, only two lines of phase 
transition meet at the multiphase point. This implies either that the
floating-paramagnetic transition line touches tangentially the 
floating-antiphase line at this point (Fig.~\ref{fig:Phase3}) or that the floating phase 
extends over the entire region from antiphase to ferromagnetic region (Fig.~\ref{fig:Phase2}).
It will be shown that a study of susceptibility and mass-gap rules out the
possibility of Fig.~\ref{fig:Phase3} and points to the validity of 
Fig.~\ref{fig:Phase2}. Thus, the final conclusion from the first perturbation
scheme is that the boundary between the ferromagnetic (antiphase) and the 
floating phase emanates from the multiphase point at an angle $\tan^{-1} -2$
($\tan^{-1} 1$) with the $\kappa$ axis.
The second perturbation scheme also supports this 
conclusion and additionally enables us to obtain an approximate phase diagram.
We shall now transform our problem to a certain problem for the nearest 
neighbour transverse Ising model and then through a study of the nearest 
neighbour model, obtain the expression for $E^{(1)}$.

\subsection {Transformation to Nearest-neighbour Model}

We start by noting that
at $\kappa=0.5$ (the `multiphase point' \cite{selke}) the ground 
state of ${\mathcal H}_{cl}$, the unperturbed Hamiltonian, is a highly 
degenerate state and any spin configuration that has no spin-domain of length
unity (i.e. no $++-++$ or $--+--$ type arrangement) can be the ground state.
The number of domain walls is immaterial and can be anything between 0 and 
$N/2$, $N$ being the total number of spins. (Of course, for periodic boundary
there can be only an even number of walls.) Let us denote the set of all such
configurations as ${\mathcal S}$. Also, let the population of this set be
$\nu$ which incidentally is of the order of $g^N$ as mentioned in Sec. I 
\cite{selke}. 
Now, the first-order correction to the eigenvalue are the eigenvalues of the 
$\nu \times \nu$ matrix $P$, whose elements are 
\[
P_{\alpha \beta} \equiv \langle \alpha | {\mathcal H}_p | \beta \rangle 
\]
where $| \alpha \rangle$ and $| \beta \rangle$ are configurations within 
${\mathcal S}$ .
The matrix $P$ can be easily given a block diagonal structure. Note that 
$s^x_j|\beta\rangle \in {\mathcal S}$ if and only if the $j$-th spin lies at 
the {\em boundary} of a domain, and the domain too has length larger than 2. 
Also, in such a case, $s^x_j$ operating on $|\beta\rangle$ will translate the 
wall at the left (right) of the $j$-th site by one lattice spacing to the
right (left). This immediately leads us to the important conclusion that 
$P_{\alpha\beta} \neq 0$ if and only if $|\alpha\rangle$ and $|\beta\rangle$
have equal number of domain walls. Thus, we can break up ${\mathcal S}$ into
subsets ${\mathcal S}(W)$, where ${\mathcal S}(W)$ contains all possible spin 
distributions with $W$ walls ($W$ = 2, 4, $\cdots N/2$). Now, the 
$\nu \times \nu$ matrix $P$ gets block-diagonalised into matrices of size 
$\nu_W \times \nu_W$,
\be
P_{\alpha \beta}(W) \equiv \langle \alpha | {\mathcal H}_p | \beta \rangle
\ee
where $\nu_W$ is the population of ${\mathcal S}(W)$ and 
$| \alpha \rangle, | \beta \rangle \in {\mathcal S}(W)$. We now observe that
the longitudinal term in ${\mathcal H}_p$ only contributes a diagonal term 
$r\Gamma(N-4W)$ to $P_{\alpha \beta}(W)$, so that one can write
\be
P(W) = M(W) + r\Gamma(N-4W){\bf 1}.
\ee
Here ${\bf 1}$ is the $\nu_W \times \nu_W$ unit matrix, 
$M_{\alpha \beta}(W) \equiv \langle \alpha | {\mathcal H}_q | \beta \rangle$
and ${\mathcal H}_q = - \Gamma \sum_j s^x_j$ is the transverse part of 
${\mathcal H}_p$. Thus the non-trivial problem is to solve the eigenproblem 
of $M(W)$.

To proceed with the matrices $M(W)$, let us construct from each member 
$| \alpha \rangle$ of ${\mathcal S}(W)$ a configuration 
$| \alpha^{\prime} \rangle$ by removing one spin from each domain. The total
number of spins in $| \alpha^{\prime} \rangle$ will obviously be
$N-W=N^{\prime}$, say. Such a transformation was also used by Villain and Bak
\cite{VB} for the case of two-dimensional ANNNI model. The set 
${\mathcal S}^{\prime}(W)$ composed of the 
states $| \alpha^{\prime} \rangle$ is then nothing but the set of all possible
distributions of $N^{\prime}$ spins with $W$ walls, with no restriction on
the domains of length unity. It is easily seen that the matrix 
\[
M^{\prime}_{\alpha \beta}(W) \equiv \langle \alpha^{\prime} | {\mathcal H}_q 
| \beta^{\prime} \rangle
\]
will then be identical with $M(W)$ since the element
\[
\langle \alpha^{\prime} | \sum_{j=1}^{N^{\prime}} s_j^x
| \beta^{\prime} \rangle
\]
is non-zero when and only when
\[
\langle \alpha | \sum_{j=1}^N s_j^x
| \beta \rangle
\]
is non-zero. The eigenproblem of $M^{\prime}(W)$
becomes simple once we we observe that ${\mathcal S}^{\prime}(W)$ is
nothing but the set of degenerate eigenstates of the Hamiltonian
\[
{\mathcal H}_0^{\prime} = - J \sum_{j=1}^{N^{\prime}} s^z_j s^z_{j+1} 
\]
corresponding to the eigenvalue
\be
E_W = - J (N^{\prime} - 2W).
\ee
Thus, if we perturb ${\mathcal H}_0^{\prime}$ by ${\mathcal H}_q$, then the 
first-order perturbation matrix will assume a block diagonal form made up of 
the matrices $M^{\prime}(W)$ for all possible values of $W$.

\subsection {Study of the Nearest-neighbour Model and Further Analysis}

To solve the perturbation 
problem for ${\mathcal H}_0^{\prime} + {\mathcal H}_q$, we note that this 
Hamiltonian is the same as ${\mathcal H}$ of Eq. (1) with $\kappa = 0$, namely,
\[ {\mathcal H}^{TI} = - \sum_{j=1}^{N^{\prime}} [ J s^z_j s^z_{j+1} + \Gamma s^x_j ].\] 
This is the Hamiltonian for the standard transverse Ising model. The exact 
solution for this Hamiltonian is readily available \cite{mattis,LSM,pf}.
The exact expression for the energy eigenstates are (for periodic chain in the
thermodynamic limit \cite{boundary}), 
\be
E= 2 \Gamma \sum_k \xi_k \Lambda_k
\ee
where $\xi_k$  may be 0, $\pm$1 and $k$ runs over $N^{\prime}/2$ equispaced 
values in the interval 0 to $\pi$. Also, $\Lambda_k$ stands for
$\surd(\lambda^2 + 2\lambda \cos k + 1)$, where $\lambda$ is the ratio 
$J/\Gamma$.  For $\Gamma=0$, the energy $E$ must be 
the same as $E_W$ of Eq.(7), so that 
\be
 2 \sum_{k=0}^{\pi} \xi_k = - (N^{\prime} - 2W).
\ee
Clearly, the different values of the quantity
\[
\left( \frac{\partial E}{\partial \Gamma}\right)_{\Gamma=0} = 
2 \sum_{k=0}^{\pi} \xi_k \cos k  
\]
correspond to the first order perturbation corrections to the different
levels. They are therefore also the eigenvalues of the $M^{\prime}$ matrix.
Thus the eigenvalues of the matrix $P(W)$ of Eq.(6) are 
\be
E_P = r\Gamma (N-4W) + 2\Gamma \sum_{k=0}^{\pi} \xi_k \cos k
\ee
Keeping $N$ fixed we have to find, for 
which value of $W$ and for which distribution of $\xi_k$, $E_P$ is
minimum subject to the constraint (9). For a given value of $\sum \xi_k$,
this minimisation is achieved if -1 values of $\xi_k$ accumulate near
$k=0$ and +1 values near $k=\pi$. Let the desired distribution be

\be \xi_k = \left\{ \begin{array}{cc}
          -1 & \mbox{for $k=0$ to $\theta$}\\
          0 & \mbox{for $k=\theta$ to $\phi$}\\
          1 & \mbox{for $k=\phi$ to $\pi$}
          \end{array}
  \right. \ee
Eq.(9) now gives 
\be N/N^{\prime} = (4\pi - \theta - \phi)/2\pi \ee 
and one obtains,
\[
E_P = - \frac{N\Gamma}{4\pi - \theta - \phi}\left[ r(4\pi - 3\theta - 3\phi)
+2(\sin \theta + \sin \phi) \right]
\]
This quantity attains a minimum value only when $\theta$ and $\phi$ are equal
and their common value ($\phi_0$, say) satisfies the condition
\be
2\pi r = \sin \phi_0 + (2\pi - \phi_0)\cos \phi_0,
\ee
The minimum value of $E_P$ is given by,
\be
E^{(1)} = - N\Gamma [ 3r - 2\cos \phi_0 ].
\ee
This is the exact expression for the first order perturbation correction
to ground state energy.  It is easily seen that for $r < - 0.5$, that is, 
for $\Gamma/J < (1-2\kappa)$, one has $\phi_0=\pi$ and $W=0$ 
(ferromagnetic phase), while for $r > 1$, that is 
$\Gamma/J < (\kappa-0.5)$, one has $\phi_0=0$ and $W=N/2$
(antiphase). As $r$ varies from $-0.5$ to 1, $\phi_0$ gradually 
changes from $\pi$ to 0 according to Eq.(13). The second derivative of
$E^{(1)}$ with respect to $r$ blows up at $r=-0.5$ and 1, indicating
two critical lines there. One can see from Eq. (13) that except for these two
values, $r$ is an analytic function of $\phi_0$ and hence, by Eq. (14), 
$E^{(1)}$ is also an analytic function of $r$. We have also checked explicitly
(Appendix A) that no higher derivative of $E^{(1)}$ with respect to $r$ blows 
up at any other value of $r$.

As mentioned in Sec. IIA, we can now conclude that the phase diagram is either
like Fig.~\ref{fig:Phase2} or like Fig.~\ref{fig:Phase3}. 

\subsection {Study of Longitudinal Susceptibility}

We shall now show that an analysis of longitudinal susceptibility points to 
the possibility of Fig.~\ref{fig:Phase2}, rather than of Fig.~\ref{fig:Phase3}.

Let us call the eigenstate of ${\mathcal H}_0^{\prime} + {\mathcal H}_q$
corresponding to $\theta=\phi=\phi_0$ as $|\psi^{\prime}\rangle$.
This state will be composed of the spin-distributions that belong to
${\mathcal S}^{\prime}(W)$ and can be written as
\[
|\psi^{\prime}\rangle = \sum_{j^{\prime}} a_{j^{\prime}} |j^{\prime}\rangle
\]
where $|j^{\prime}\rangle$ runs over all the states in 
${\mathcal S}^{\prime}(W)$. If we construct from each state 
$|j^{\prime}\rangle$ another state $|j\rangle$ by adding one spin to each
domain, and then combine these states with the same coefficients, then
we arrive at a state $\sum_j a_j |j\rangle$ where $a_j=a_{j^{\prime}}$. This 
is an eigenstate of $M(W)$ and hence
of $P(W)$ (see Eq. (6)) and this eigenstate is nothing but the zero-th order
eigenfunction $|\psi^{(0)} \rangle$ for the perturbed ground state of
${\mathcal H}_{cl} + {\mathcal H}_p$. One should observe that although
the spin-spin correlation may not be equal for $|j\rangle$ and
$|j^{\prime}\rangle$, the longitudinal magnetisation $M_z$ must be
the same for them (- equal number of positive and negative spins are added
while transforming $|j\rangle$ to $|j^{\prime}\rangle$). Thus, the
longitudinal susceptibility
\[
\chi_z \propto \langle M_z^2 \rangle - \langle M_z\rangle^2
\]
of $|\psi^{(0)} \rangle$ must be the same as that of $|\psi^{\prime}\rangle$.
The spin-spin correlation 
\[
C^z(n) \equiv \langle s_i^z s_{i+n}^z \rangle
\]
for $|\psi^{\prime}\rangle$ may be calculated (see Appendix B). 
In the case of $\Gamma < J$, for the entire range 
$0 < \phi_0 < \pi$, the correlation is
\[ C^z(n) = A \frac{1}{\sqrt n}\cos [n(\pi - \phi_0)] \]
where $A$ is a constant. This is clearly a floating phase. The
susceptibility $\chi_z$ is hence infinity for both the states 
$|\psi^{\prime}\rangle$ and $|\psi^{(0)} \rangle$. This leads us to the
conclusion that the zero-th order eigenstate is in floating phase and hence, at 
least for small values of $\Gamma$, the ground state of transverse ANNNI chain
must be a floating phase for all values of $r$ between $-0.5$ and 1. Of course, 
for large values of $\Gamma$ the perturbation corrections may cancel the 
divergence of susceptibility and lead to a paramagnetic state.

One must note that it is difficult to derive an {\em exact} relationship
between the correlation in state $|\psi^{(0)} \rangle$ and the same in state
$|\psi^{\prime}\rangle$. Although a similar relationship was obtained by 
Villain and Bak \cite{VB}, we do not extend that derivation here. It is
interesting to note that Villain and Bak (and also Uimin and Rieger 
\cite{rieger}) assumed the wave number to be equal
to the number of domain walls per site, but we shall soon find that this
conclusion disagrees with the results obtained from our second perturbation 
scheme.

\subsection {Study of Mass Gap}

We shall now show that an analysis of mass gap also points to the possibility 
of Fig.~\ref{fig:Phase2}, rather than of Fig.~\ref{fig:Phase3}, thus agreeing 
with the conclusion from the analysis of longitudinal susceptibility.

One signature of floating phase or diverging correlation length is vanishing 
{\em mass-gap} \cite{mattis,sachdev}. Let us now study the first order 
(in $\Gamma$) correction to the mass-gap. Since the first order 
correction to all energy states is given by the different possible values of
$E_P$ of Eq. (10), the correction to the energy of the first excited state 
is the smallest
possible value of $E_P$ apart from the ground state $E^{(1)}$. To find the
lowest excitation over the ground state, we note that such excitation is
possible either (i) by keeping $\sum \xi_k$ fixed and rearranging the $\xi_k$
values; or (ii) by altering $\theta$ and $\phi$ and thus altering $\sum \xi_k$.
For (i) the lowest excitation will correspond to an interchange of +1 and -1
at $k=\phi_0$, which will lead to a mass gap (for the whole system, not per
site)
\[ \Delta^{(1)} = \frac{8\pi\Gamma\lambda \sin \phi_0}
{N^{\prime}\Lambda_{\phi_0}}.  \]
For (ii) the mass gap is :
\[ \Delta^{(1)} = \frac{1}{2} \left(\frac{\partial^2 E_P}{\partial \theta^2}
\right)_{\theta=\phi_0}(\delta \theta)^2 \]
Here $\delta \theta$ is the smallest possible deviation in $\theta$ at 
$\phi_0$. As the smallest possible change in $W$, and hence in $N^{\prime}$ 
is 2, Eq.(12) (with $\theta=\phi$) tells us that 
\[ \delta \theta = \frac{2(2\pi - \theta)^2}{\pi N} \sim \frac{1}{N}. \]
This shows that for both the mechanisms (i) and (ii), the mass gap 
$\Delta^{(1)}$ vanishes as $N \rightarrow \infty$ for
all values of $\phi_0$ between 0 and $\pi$. This shows that for all values 
of $r$ between $-0.5$ and 1 there must be floating phase for small $\Gamma$.

For transverse Ising model ($\kappa=0$) the phase transition occurs at 
$\Gamma=J$ and at $\Gamma=J + \epsilon$
the first order (in $\epsilon$) correction to the mass gap is just $\Gamma$,
and is thus nonvanishing, indicating that the divergent correlation length
does not extend beyond $\Gamma=J$.

\section{The Second Perturbation Scheme}

\subsection {Principle}

In the previous section we have found that the phase diagram looks like
Fig.~\ref{fig:Phase2}. In order to reconfirm this result explicitly, we have 
to know the eigenfunction with first- or even zero-th order 
correction. As it is difficult to calculate the eigenfunction for the 
perturbation scheme discussed above in Sec. II, we shall perform the second 
set of perturbation calculation now. 

Let us write the Hamiltonian ${\mathcal H}$ of Eq. (1) for $\kappa < 0.5$ as
\be
{\mathcal H} = N (h_0 + h_1 - J \kappa )
\ee
with
\be
h_0 = - (1-2\kappa) J \frac{1}{N} \sum_j  s^z_j s^z_{j+1} -
\Gamma \frac{1}{N} \sum_j s^x_j
\ee
and 
\be
h_1 = \kappa J  \frac{1}{N} \sum_j (1-s^z_j s^z_{j+1})(1-s^z_{j+1} s^z_{j+2}).
\ee
Clearly, $h_0$ represents the Hamiltonian of standard transverse Ising
model with nearest-neighbour interaction and $h_1$ represents the number
of domains of length 1. We shall treat the operator $h_1$ as perturbation on
the Hamiltonian $h_0$.
All the eigenstates of $h_0$ are precisely known \cite{pf} and 
for each eigenstate one can readily calculate the expectation value of
$h_0 + h_1$ and identify the eigenstate for which $<h_0 + h_1>$ is the lowest.
This state is the eigenstate with zero-th order perturbation correction. 
Therefore, the basic idea is to identify the ground state upto first order 
perturbation correction to eigenvalue. We carry on this scheme for all values
of $\kappa$ ($< 0.5$) and $\Gamma$ and in each case calculate the correlation 
function (as defined in Eq. (2)) and characterise the phase therefrom. 

Indeed, our results are reliable only for small values of $<h_1>$. Hence, 
this quantity must be small
for the low-lying eigenstates of $h_0$. As mentioned already, the operator
$h_1$ basically counts the number of domains of length 1. For $\kappa < 0.5$ 
the ground state of $h_0$ at $\Gamma=0$ is ferromagnetic and the low-energy 
states may be expected to have small $<h_1>$ and the perturbation treatment 
will be justified for small values of $\Gamma$. However, for $\kappa > 0.5$, 
the ground state of $h_0$ will be 
antiferromagnetic (having domains of length 1 only) and such perturbation 
scheme is not valid. 

For $\kappa > 0.5$ one can break up ${\mathcal H}$ as 
\be {\mathcal H} = N(h^{\prime}_0 + h^{\prime}_1 - 0.5 J) \ee
with the unperturbed Hamiltonian defined as 
\be
h^{\prime}_0 = (\kappa - 0.5) J \frac{1}{N} \sum_j  s^z_j s^z_{j+2} - 
\Gamma \frac{1}{N} \sum_j s^x_j
\ee
and the perturbation as
\be
h^{\prime}_1 = 0.5 J \frac{1}{N} \sum_j (1-s^z_j s^z_{j+1})(1-s^z_{j+1} s^z_{j+2})
\ee
The Hamiltonian $h^{\prime}_0$ represents a chain with only next-nearest 
neighbour interaction and the operator $h^{\prime}_1$ counts (as for 
$\kappa < 0.5$) the number of domains of length 1.
The ground state of $h^{\prime}_0$ at $\Gamma=0$ is antiphase, and hence has
$<h^{\prime}_1> = 0$, so that the perturbation scheme is justified. Following
the principle for the previous case, one can identify numerically the 
eigenstate (of $h^{\prime}_0$) for which $<h^{\prime}_0 + h^{\prime}_1>$ is a 
minimum. From the correlation function of this state, one can characterise the 
phase.

The phase diagram obtained from the analysis of the present section is shown
in Fig.~\ref{fig:boundary}. It has the following features : (i) The floating phase exists and 
extends from the ferromagnetic phase to the antiphase for small $\Gamma$; thus
Fig.~\ref{fig:Phase2} rather than Fig.~\ref{fig:Phase3} is the type of the phase diagram. (ii) There is
a line along which transition from floating phase to paramagnetic phase takes
place; this line for $\kappa < 0.5$ (obtained using Eq.(15)) continues smoothly
to the same for $\kappa > 0.5$ (obtained using Eq.(18)).

We shall now present the details of the calculation for the two ranges of 
$\kappa$.

\subsection {For $\kappa < 0.5$}

The Hamiltonian $h_0$ of Eq.(16) corresponds to nearest-neighbour transverse
Ising model with (ferromagnetic) interaction strength $(1-2\kappa)J$. 
Following Eq.(8), the eigenstates of this operator are 
\be
< h_0> =  \Gamma \frac{1}{N/2} \sum_k \xi_k \Lambda_k
\ee
with $\Lambda_k = \surd(\lambda^2 + 2\lambda \cos k + 1)$ as in Sec. II, but
$\lambda=(1-2\kappa)J/\Gamma$. The line corresponding to a 
constant value of $\lambda=$ is now a straight 
line at an angle $\tan^{-1}(2/\lambda)$ with the $\kappa$ axis (Fig.~\ref{fig:new1}).
For each eigenstate, the $n$-th neighbour
longitudinal correlation function (defined by Eq. (2)) can be expressed as 
the Toeplitz determinant \cite{pf}
\be
  C^z(n) = \left | \begin {array}{ccccc}
    G_0 & G_{-1} & G_{-2} & \cdots & G_{-n+1} \\
    G_1 & G_0 & G_{-1} & \cdots & G_{-n+2} \\
    G_2 & G_1 & G_0 & \cdots & G_{-n+3} \\
    \cdots \cdots & & & & \\
    G_{n-1} & G_{n-2} & G_{n-3} & \cdots & G_0 \\
  \end{array}      \right |
\ee
where the elements are given by
\be G_j = - \frac{2}{N} \sum_{k=0}^{\pi} 
\frac{\xi_k}{\Lambda_k}[\cos (kj-k) + \lambda \cos(kj) ] \ee
Here, the wave vector $k$ runs over the $N/2$ equispaced values in the interval
0 to $\pi$ and $\xi_k$ can be 0 or $\pm 1$ for each of the $N/2$ values of $k$
between 0 and $\pi$.)

Now, for any eigenstate of $h_0$, the expectation value of
$h_1$ can be calculated first by rewriting it as,
\[
h_1 = \kappa J \frac{1}{N} \sum_j (1-2s^z_j s^z_{j+1}+ s^z_j s^z_{j+2})
\]
and then evaluating the first and second neighbour correlation functions using
Eq.(22). The result is,
\be
< h_1 > =  \kappa J [(1-G_0)^2 - G_1 G_{-1}].
\ee
To find which distribution of $\xi_k$ gives 
the smallest value of $<h_0 + h_1>$, we note that $G_0$
will be closest to 1, when the $(-1)$ values of $\xi_k$ accumulate near $k=0$ 
and $(+1)$ values near $k=\pi$. Although Eq.(24) does not clearly indicate 
that such a distribution of $\xi_k$ will also lead to the largest values of
$G_1$ and $G_{-1}$, we have verified by going through {\em all} the $2^N$
states (for $N=12$) that indeed such a distribution leads to the lowest value
of $<h_0 + h_1>$. We can now assume the distribution of $\xi_k$ values to be the
same as described in Eq. (11) and calculate $<h_0 + h_1>$ for given values of
$\theta$ and $\phi$. For every $\kappa$ and $\Gamma$, we take a 
system of 1000 spins and consider all
possible choices of $\theta$, $\phi$ and note the values (say, $\theta_0$ 
and $\phi_0$) for which $<h_0 + h_1>$ attains a minimum. For the entire 
parameter range investigated in this work, we found $\theta_0 = \phi_0$. 
Eq.(11) now reduces to (with $\theta =\phi=\phi_0$) a distribution, namely
\be \xi_k = \left\{ \begin{array}{cc}
        -1 & \mbox{for $k=0$ to $\phi_0$}\\
        1 & \mbox{for $k=\phi_0$ to $\pi$.}
        \end{array}
\right. \ee
This corresponds to our approximate ground-state eigenfunction for
${\mathcal H}$. We have also found that for this
eigenfunction, $<h_1>$ is non-zero. As $<h_1>$ is zero for all the ground state
eigenfunctions at $\Gamma=0$, we may conclude that the state we have determined
as the ground state of $h_0 + h_1$, is an excited state of 
$h_0$. In this connection, we point out that for 2D ANNNI model, it has been 
proved that \cite{SDG} if one neglects completely the domains of length 1 
(i.e. assumes $<h_1>$ to be zero), then one misses the floating phase,
indicating that the floating phase consists of states that do {\em not} belong
to the ground state at the $(T,\kappa) = (0,0.5)$ point.

For every value of $\kappa$ and $\Gamma$ we now have determined the value
of $\phi_0$ and Eq.(25) therefore gives us the ground state eigenfunction.
This leads to an expression for the correlation function $C^z(n)$ of Eq.(22) 
in the form of a determinant, an analytic expression of which has
been given in Appendix B for large values of $n$. The final result is as 
follows. For $\lambda > 1$, that is, $\Gamma/J < (1-2\kappa)$, the value of
$\phi_0$ turns out to be $\pi$, leading to a ferromagnetic phase. For 
$\lambda <  1$, that is, $\Gamma/J > (1-2\kappa)$, there are two regions. In
one region $\phi_0 < \pi$ and according to Appendix B, the correlation is
\[ C^z(n) = A(\phi_0) \frac{1}{\sqrt n} \]
indicating a floating phase with index 0.5 and no modulation. The parameter
$\phi_0$ varies continuously as a function of $\kappa$ and $\Gamma$ 
(Fig.~\ref{fig:klt5.phi0}), 
but this variation affects only the amplitude $A$. In the other
region, $\phi_0 = \pi$, and the correlation is exponentially decaying, 
indicating a paramagnetic phase. The values of $\lambda$ and $\phi_0$ and the 
corresponding phases in different portions of the phase diagram is indicated 
in Fig.~\ref{fig:new2}. 
The phase diagram thus obtained agrees with
the conclusions from our first perturbation scheme and is presented in Fig.~\ref{fig:boundary}.

\subsection {For $\kappa > 0.5$}

The Hamiltonian $h^{\prime}_0$ of Eq.(18) corresponds to a chain 
(${\mathcal C}$, say) with only next-nearest neighbour interaction and can be 
broken into two independent transverse Ising chains (${\mathcal C}_1$ and 
${\mathcal C}_2$, say) each having nearest-neighbour antiferromagnetic 
interaction. Hence, the correlation $\sum s^z_j s^z_{j+1}$ 
in ${\mathcal C}$ will be zero and $\sum_j s^z_j s^z_{j+2}$ in ${\mathcal C}$ 
will be the same as the nearest-neighbour correlation in the constituent 
chains ${\mathcal C}_1$ and ${\mathcal C}_2$. (Indeed, ${\mathcal C}_1$ and
${\mathcal C}_2$ will be uncorrelated because, any configuration
of ${\mathcal C}_1$ will couple with equal probability to two configurations
of ${\mathcal C}_2$, one of which can be obtained from the other by reversing
all spins.) Moreover, since each of the 
antiferromagnetic chains can be transformed to a ferromagnetic chain 
(${\mathcal C}_0$, say) by simply reversing the alternate spins, the 
eigenstates of $h_0^{\prime}$ in ${\mathcal C}$ are related to the 
eigenstates of a nearest-neighbour ferromagnetic transverse Ising chain 
${\mathcal C}_0$. We have then,
\be
< h^{\prime}_0> =  \Gamma \frac{1}{N/2} \sum_k \xi_k \Lambda_k
\ee
and
\be
< h^{\prime}_1 > =   0.5 J (1-G_0)
\ee
with $G_0$ defined by Eq. (23) but now with $\lambda=(\kappa-0.5)J/\Gamma$. 
As before, $k$ runs over $N/2$ equispaced values in the interval $0$ to $\pi$.
The $\lambda=$ constant line is now a straight 
line at an angle $\tan^{-1}(1/\lambda)$ with the $\kappa$ axis (Fig.~\ref{fig:new1}).
As in the previous subsection, we look for the eigenstate, (i.e. the $\xi_k$ 
distribution) for which $< h^{\prime}_0 + h^{\prime}_1 >$ will be lowest and
observe that $G_0$ will be largest when the $(-1)$ values of $\xi_k$ 
accumulate near $k=0$ and $(+1)$ values near $k=\pi$. Also, investigating all
the states for a chain of only 12 spins, we find that the minimum value of 
$< h^{\prime}_0 + h^{\prime}_1 >$ corresponds
to the distribution with no zero values of $\xi_k$. Thus, assuming the
distribution of Eq. (25), 
we can find numerically the value of $\theta$  (say, $\phi_0$) for which 
$< h^{\prime}_0 + h^{\prime}_1 >$ will be lowest for a chain of 1000 spins.
It is found that this minimum correponds to a non-zero value of 
$<h^{\prime}_1>$,
indicating that as for $\kappa < 0.5$, the unperturbed state corresponding to 
the ground state is an excited state of $h^{\prime}_0$.

\begin{figure}
\noindent \includegraphics[clip,width= 6cm, angle = 270]{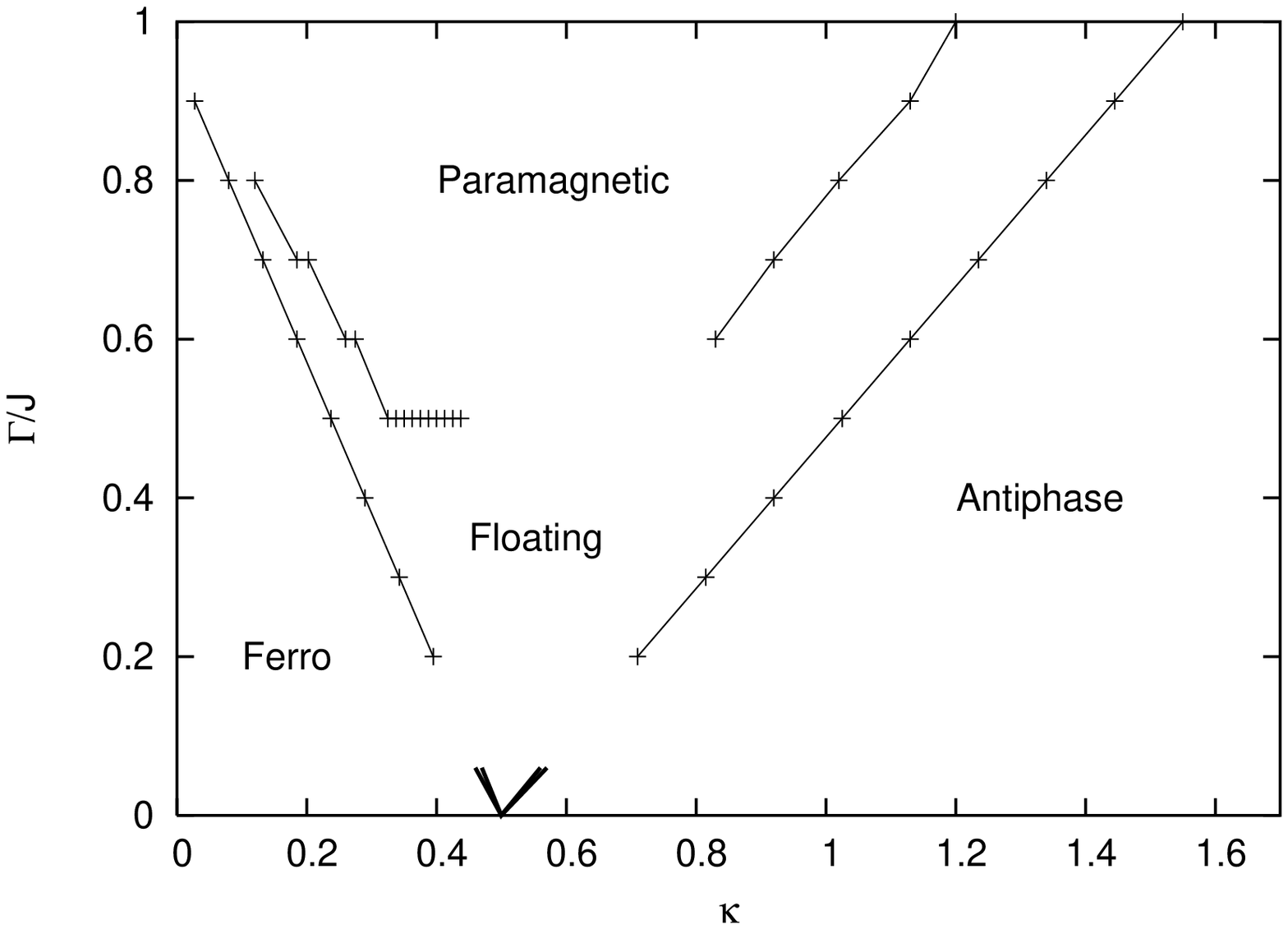}
\caption\protect{\label{fig:boundary}Phase diagram of the transverse ANNNI 
model as obtained from the first and the second perturbation scheme. 
The thick lines are exact results from the first perturbation scheme. The thin
lines are approximate results from the second perturbation scheme. The thin 
line is not drawn in the region where the latter scheme becomes unreliable.}
\end{figure}

\begin{figure}
\noindent \includegraphics[width= 8cm, angle = 0]{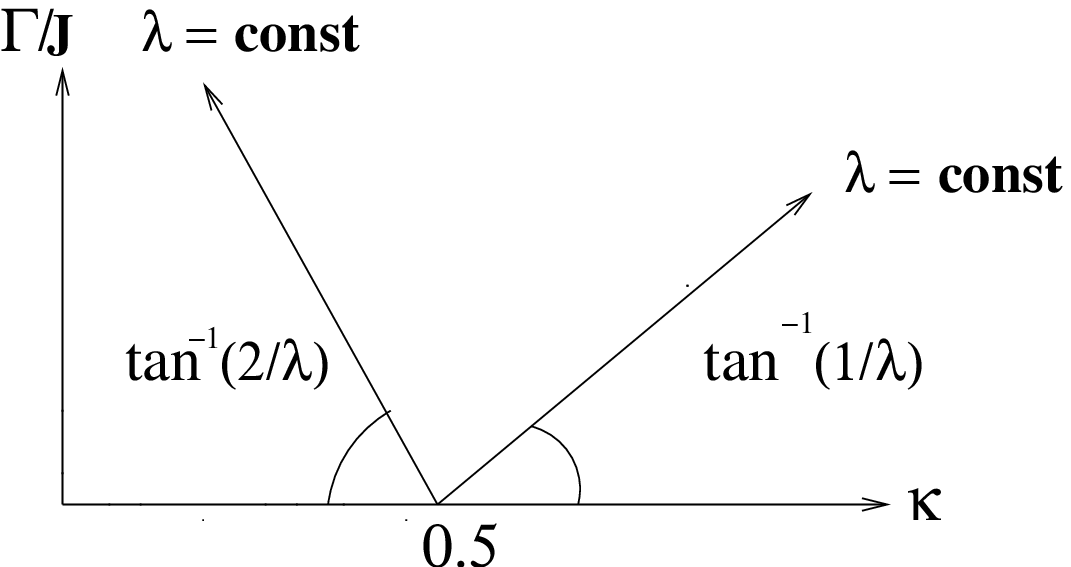}
\caption\protect{\label{fig:new1} Plot of $\lambda=$ constant line with the $\kappa$ axis.}
\end{figure}

\begin{figure}
\noindent \includegraphics[clip,width= 6cm, angle = 270]{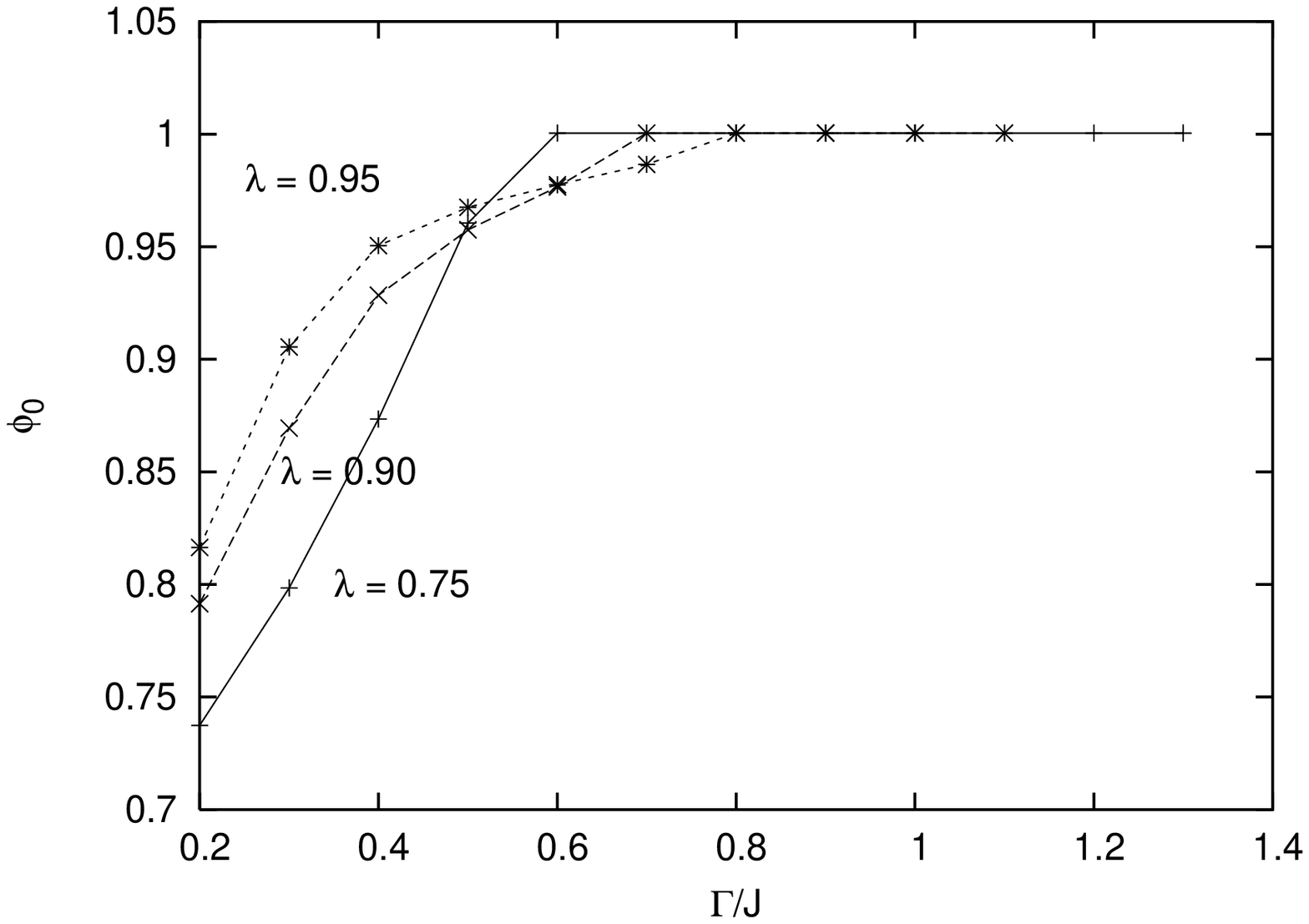}
\caption\protect{\label{fig:klt5.phi0}Plot of $\phi_0$ vs. $\Gamma$ 
for $\kappa < 0.5$. Here $\lambda=(1-2\kappa)J/\Gamma$.}
\end{figure}

\begin{figure}
\noindent \includegraphics[clip,width= 6cm, angle = 0]{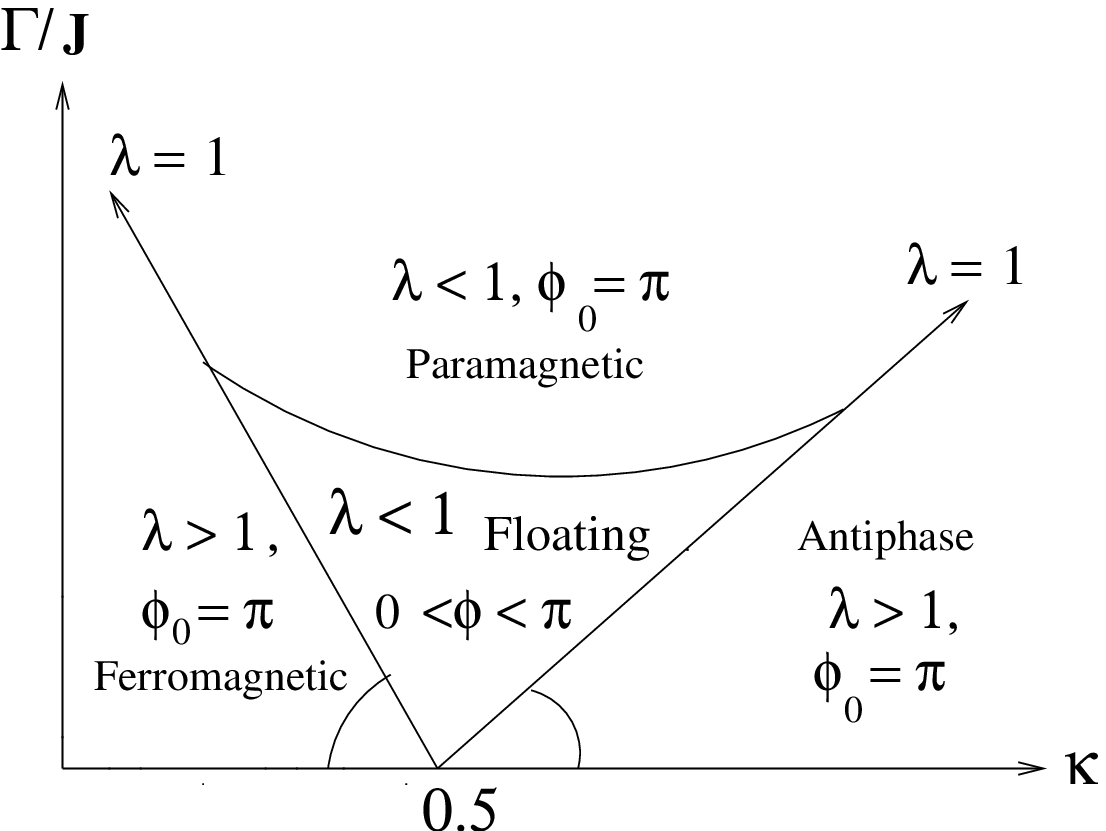}
\caption\protect{\label{fig:new2} Schematic phase diagram showing the values of
$\lambda$ and $\phi_0$ in different portions.}
\end{figure}

\begin{figure}
\noindent \includegraphics[clip,width= 6cm, angle = 270]{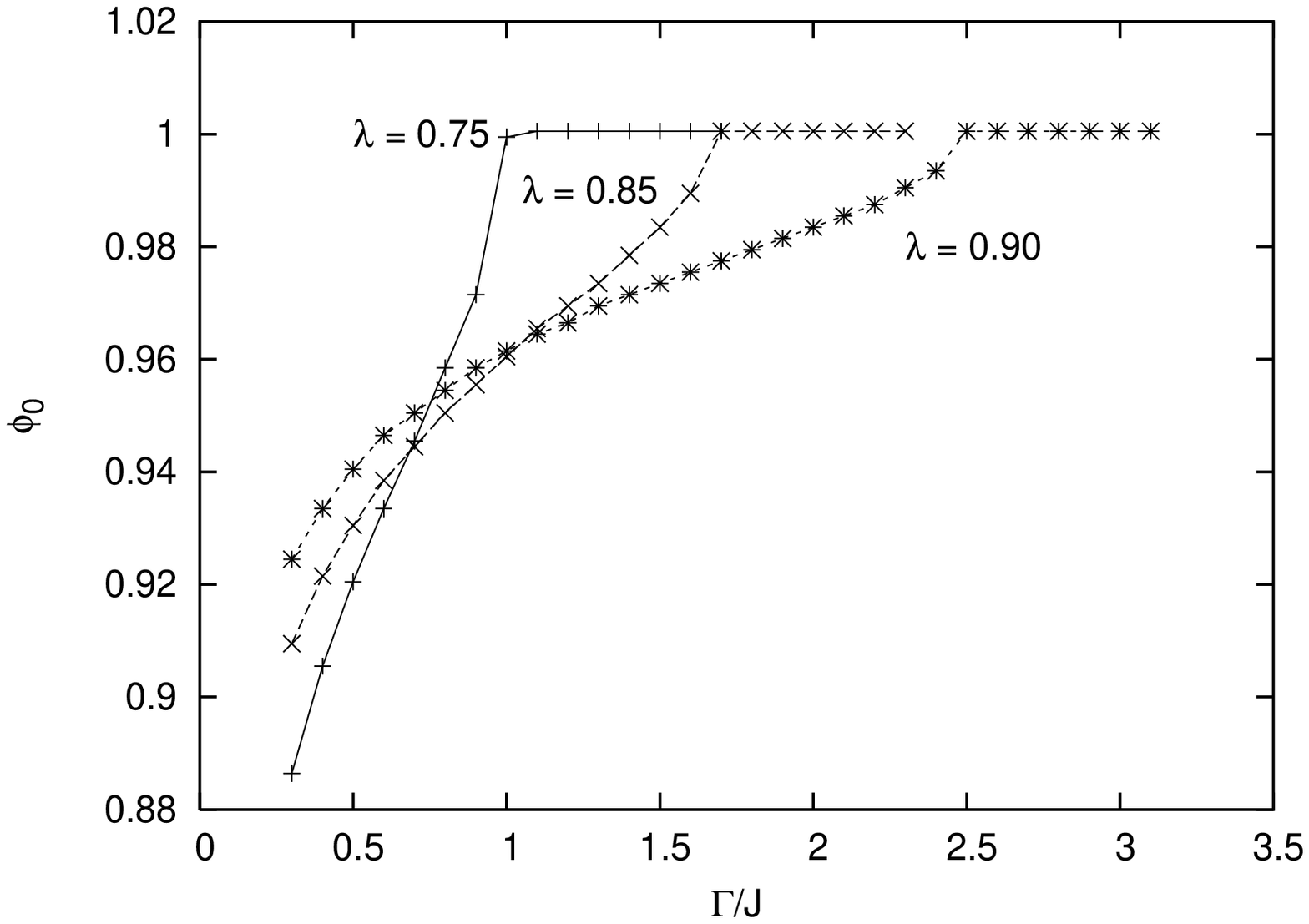}
\caption\protect{\label{fig:kgt5.phi0}Plot of $\phi_0$ vs. $\Gamma$
for $\kappa > 0.5$. Here $\lambda=(\kappa-0.5)J/\Gamma$}
\end{figure}

As in the previous subsection, we can use the expression for $C^z(n)$ 
mentioned in the Appendix B as the correlation in the chain ${\mathcal C}_0$. 
The corresponding spin-spin correlation in the ANNNI chain ${\mathcal C}$ will
be 
\[ C^z_A(n) = \left\{ \begin{array}{ll}
        0 & \mbox{for odd $n$}\\
        (-1)^n C^z(n/2) & \mbox{for even $n$}
        \end{array}
\right. \]
In this expression, the $(-1)^n$ factor takes account of the mapping between 
the ferromagnetic and antiferromagnetic chains. The final result is as follows. 
For $\lambda > 1$, that is, $\Gamma/J < (\kappa - 0.5)$, the value of
$\phi_0$ turns out to be $\pi$, so that the correlation in the chain 
${\mathcal C}_0$ is constant and the correlation $C^z_A(n)$
corresponds to a perfect antiphase. For
$\lambda <  1$, that is, $\Gamma/J > (\kappa - 0.5)$, there are two regions. In
one region $\phi_0 < \pi$ and according to Appendix B, the correlation is
\be C^z_A(n) = A(\phi_0) \cos\left(\frac{n\pi}{2}\right)\frac{1}{\sqrt n} \ee
indicating a floating phase with index 0.5 and modulation
\[ q = \pi/2 .\] 
The parameter 
$\phi_0$ varies continuously as a function of $\kappa$ and $\Gamma$ 
(Fig.~\ref{fig:kgt5.phi0}),
but this variation affects only the amplitude $A$. 
In the other region, $\phi_0 = \pi$, and the correlation in ${\mathcal C}_0$
is exponentially decaying, indicating a paramagnetic phase with the same 
modulation. The values of $\lambda$ and $\phi_0$ and the 
corresponding phases in different portions of the phase diagram is indicated 
in Fig.~\ref{fig:new2} and the resulting phase diagram is presented in 
Fig.~\ref{fig:boundary}, which agrees with the conclusions from our first 
perturbation scheme.

\section{Discussion}

In this section we shall point out some features of the phase diagram we have
obtained. 

(1) In the floating phase ($\lambda < 1$) the spin-spin correlation decays as
\[ C^z(n) = A\frac{1}{n^{\eta}} \]
with $\eta$ = 0.5. It is interesting to note that on the boundary between the 
floating and the ordered (ferromagnetic or antiphase) state $\lambda$ is 1 
and $\phi_0$ is 0 or $\pi$ and the correlation decays algebraically with 
$\eta$ = 0.25 as shown by Pfeuty \cite{pf}. Thus, the value of the index
undergoes a non-analytic change at the boundary.

\begin{figure}
\noindent \includegraphics[clip,width= 6cm, angle = 270]{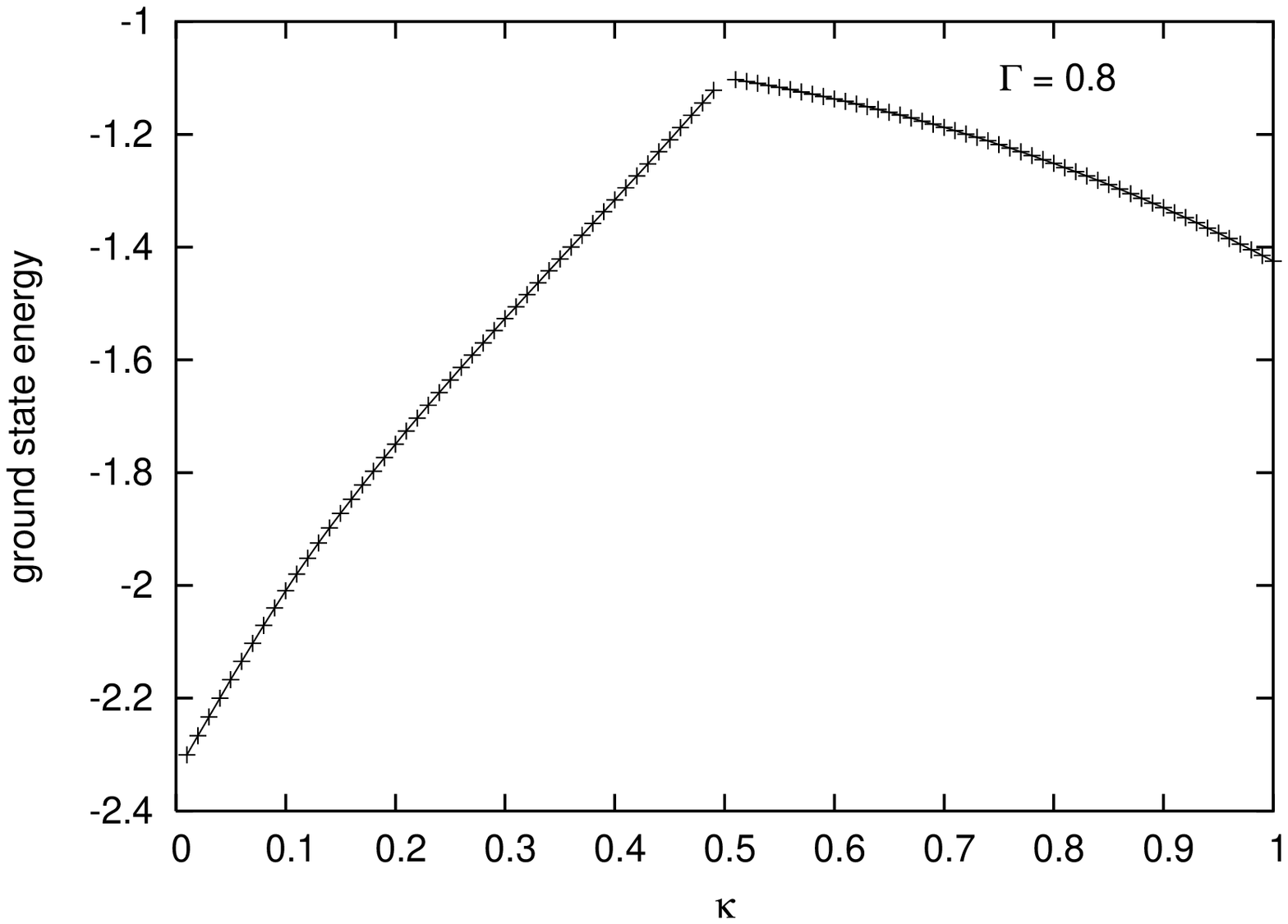}
\caption\protect{\label{fig:ground}Plot of ground state energy vs.
$\kappa$ for $\Gamma = 0.8$}
\end{figure}

(2) In the floating phase, the correlation is non-oscillatory ($q=0$) for 
$\kappa < 0.5$ and oscillatory ($q=\pi/2$) for $\kappa > 0.5$. Hence, the 
line $\kappa=0.5$ is the ``disorder line'' \cite{selke} across which the 
modulation wavevector suffers a sudden change from 0 to $\pi/2$. We have 
also measured the perturbed energy $<h_0>+<h_1>$ and 
$< h^{\prime}_0>+< h^{\prime}_1>$ on two sides of the disorder line for a 
given $\Gamma$ (Fig.~\ref{fig:ground}). This energy remains continuous but suffers a change
of slope at $\kappa = 0.5$. Since the first perturbation scheme does not
indicate any phase transition around $\kappa = 0.5$, we guess that this 
change of slope is not the signature of any serious non-analytic behaviour 
but is only an outcome of the approximation inherent in perturbation 
calculation.

\begin{figure}
\noindent \includegraphics[clip,width= 6cm, angle = 270]{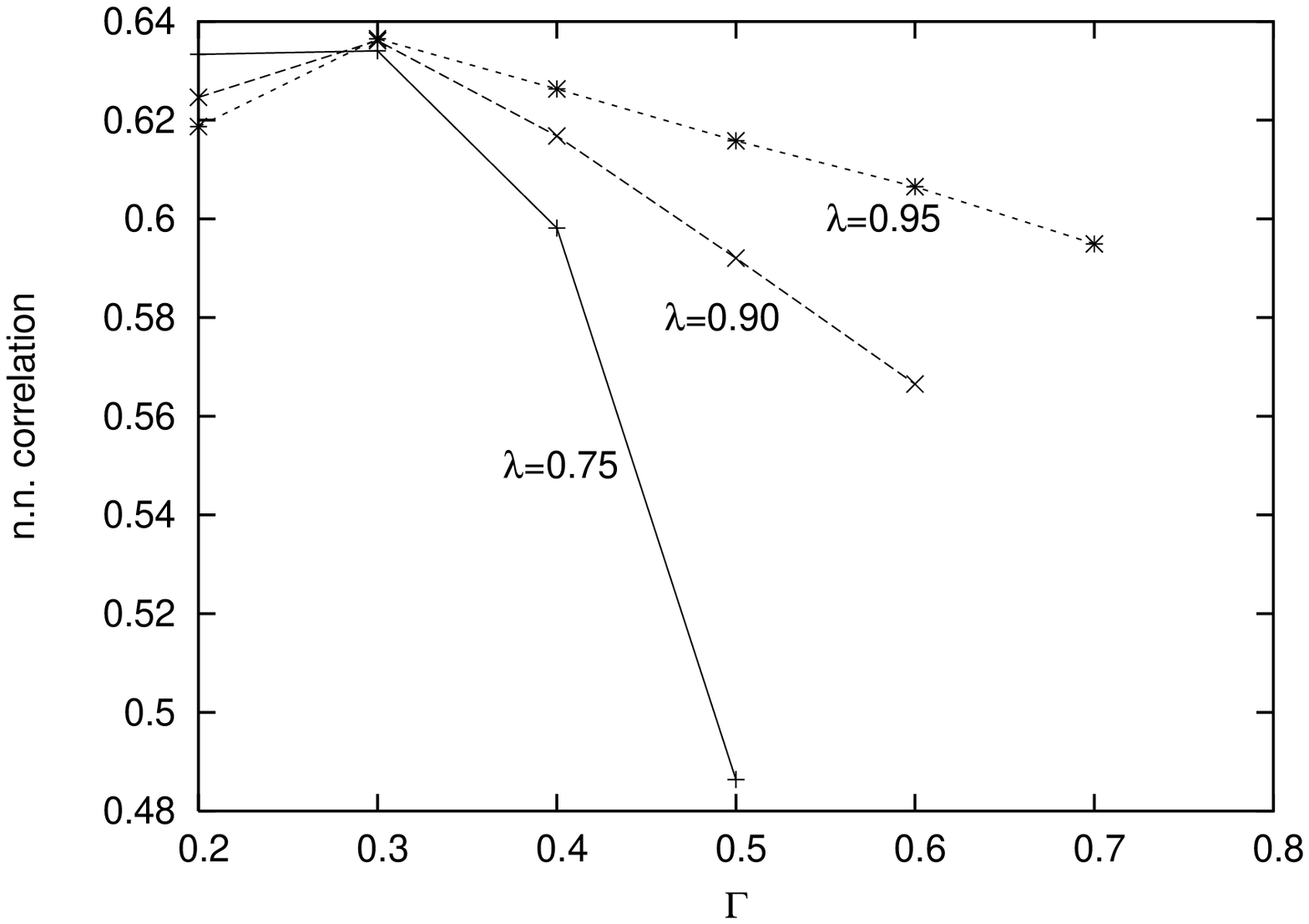}
\caption\protect{\label{fig:nncorr}Plot of nearest neighbour correlation as a
function of $\Gamma$ for $\lambda$=$0.75, 0.90$ and $0.95$}
\end{figure}

(3) Within the floating phase, for $\kappa > 0.5$, the nearest-neighbour 
correlation is zero (Eq. (28)) and the wave-vector is $q=\pi/2$, implying that
the wave number $q/2\pi$ is equal to the number of domain walls per site.
On the other hand, for $\kappa < 0.5$, the nearest-neighbour correlation
is not at all zero (Fig.~\ref{fig:nncorr}), but the correlation is non-oscillatory ($q=0$),
implying that the wave number $q/2\pi$ is {\em not} equal to the number of 
domain walls per site. One must note that (as mentioned above), the analytic
treatments of Villain and Bak \cite{VB}, and of Uimin and Rieger \cite{rieger}
assumes the equality of the wave-number and the number of domain walls per 
site. This assumption thus agrees with our results for $\kappa > 0.5$ but not
for $\kappa <  0.5$.

(4) The second perturbative scheme would be reliable only when the ratio 
$<h_1>/<h_0>$ or $< h^{\prime}_1>/< h^{\prime}_0>$
(accordingly as $\kappa$ is $< 0.5$ or $> 0.5$) is small. We have included in 
Fig.~\ref{fig:boundary} only those points where this ratio is less than some
arbitrarily chosen number 1/3. This criterion made us unable to locate the 
boundary between the paramagnetic and floating phase near $\kappa = 0.5$,
$\Gamma = 0$
resulting in a gap there in Fig.~\ref{fig:boundary}. Some better 
approximation is needed to bridge this gap.

(5) The first and second perturbation schemes can be compared and reconciled 
in the following way. For the first perturbation scheme, the unperturbed 
Hamiltonian corresponds to the $\kappa=0.5$, $\Gamma=0$ point, and the exact 
expressions for the first order correction to energy gives exactly the 
directions along which the boundaries of ferromagnetic phase and antiphase 
emanate from this point (see Fig. 5). On the other hand, for the second 
perturbation scheme, the unperturbed Hamiltonian corresponds to the boundary 
lines of the ferromagnetic phase and the antiphase since on these lines
$<h_1>$ and $<h^{\prime}_1>$ are zero. The approximate estimate
of the first order correction to energy gives the nature of the phase near the
boundary. However, since the estimate is only approximate, at some regions of
the parameter space, it is not reliable (as explained above). 

\begin{acknowledgements}
It is a pleasure to acknowledge the encouragement and fruitful discussions with
P. Sen, M. Barma and B. K. Chakrabarti. The possibility of Fig. 4 was pointed
out by M. Saha. The work of one author (AKC) was supported by UGC fellowship. 
We acknowledge the financial support from DST-FIST for computational facility.\\
\end{acknowledgements}

{\bf \begin{center}
\appendix{APPENDIX A \\ 
HIGHER DERIVATIVES FOR THE FIRST ORDER PERTURBATION CORRECTION} \\
\end{center} } 

We have derived earlier an expression 
\[ E^{(1)} = - N\Gamma\frac{ 2\sin \phi_0 + 2\pi r - 3 r \phi_0 }{2\pi - \phi_0}
\]
for the first order correction to the ground state energy of the Hamiltonian
(see Eq. (3))
\[
{\mathcal H}_{cl} = - J\sum_j [ s^z_j s^z_{j+1} - \frac{1}{2} s^z_j s^z_{j+2}]
\]
under perturbation by the Hamiltonian (see Eq. (4))
\[
{\mathcal H}_p = \sum_j \Gamma \left[ -s^x_j + r s^z_j s^z_{j+2} \right].
\] 
In this Appendix 
we shall show by the method of induction, that for all values of $\phi_0$ in 
the range
$0 < \phi_0 < \pi$, all higher derivatives of this quantity $E^{(1)}$ remain 
finite. For this purpose, we note that the first and second order derivatives 
are :
\[ \frac{\partial E^{(1)}}{\partial r} = -N\Gamma\left( -\frac{4\pi}{(2\pi-\phi_0)} 
+ 3 \right) \]
and
\[ \frac{\partial^2 E^{(1)}}{\partial r^2} = -N\Gamma\left( \frac{8\pi^2}
{(2\pi-\phi_0)^{3}\sin \phi_0} \right). \]
Here we have used the equality 
\[ \frac{\partial \phi_0}{\partial r} = \frac{2\pi}{(2\pi-\phi_0)\sin \phi_0} \]
which can be obtained from Eq. (13).
Thus, at least for $m=1$ and 2 the derivatives can be expressed in the form 
\be \frac{\partial^m E^{(1)}}{\partial r^m} = -N\Gamma\sum_{i=1}^{m^{\prime}} 
k_i\frac{\cos^{\alpha_i}
{\phi_0}}{{(2\pi-\phi_0)^{\beta_i}}{\sin^{\gamma_i}{\phi_0}}} \ee 
where $m^{\prime} \sim 3^m$, and $\alpha_i$, $\beta_i$, $\gamma_i$, $k_i$ are 
constants $\ge 0$ with the restriction $\gamma_{i} \le \alpha_i$ (except 
for $m = 1$, where $\alpha_i = \gamma_i = 0 $). Differentiating Eq.(29) once,
it can be easily verified that the form (29) should be valid for $m+1$ also,
and is hence valid for all values of $m$.
Since the right hand side of Eq. (29) diverges only when $\sin \phi_0$ 
vanishes, we conclude that so long as $\phi_0$ is neither 0 or $\pi$
the higher derivatives of $E^{(1)}$ remain finite. 
\\

{\bf \begin{center}
\appendix{APPENDIX B \\
LONG RANGE CORRELATION FOR THE EXCITED STATES} \\
\end{center} }

Here we shall consider transverse Ising model with nearest neighbour 
interaction described by the Hamiltonian
\be {\mathcal H}^{TI} = - \sum_{j=1}^N [ J s^z_j s^z_{j+1} + \Gamma s^x_j ].\ee
The exact solution \cite{mattis,LSM,pf} for this Hamiltonian tells us that the
$2^N$ number of energy eigenvalues are 
\be E= 2 \Gamma \sum_k \xi_k \Lambda_k \ee
where $\Lambda_k=\surd(\lambda^2 + 2\lambda \cos k + 1)$, $\lambda=J/\Gamma$,
$\xi_k$  may be 0, $\pm$1 and $k$ runs over $N/2$ equispaced values in the 
interval 0 to $\pi$. 

In the text we have come across (more than once) $\xi_k$ distribution of the 
following type :
\be \xi_k = \left\{ \begin{array}{cc}
        -1 & \mbox{for $k=0$ to $\phi$ (unexcited)}\\
        1 & \mbox{for $k=\phi$ to $\pi$ (excited)}
        \end{array}
\right. \ee
This is an excited state and reduces to the ground state only for $\phi = \pi$.
This state is important for this work because it corresponds to the ground 
state after perturbation in the second perturbation scheme (Eq. (25)). It also
corresponds to the first order correction in the first perturbation scheme 
(Eq. (11)).
We shall present in this Appendix the expressions for the longitudinal 
two-spin correlation function
\be C^z(n) \equiv < s^z_j s^z_{j+n} > - <s^z_j>^2 \ee
(for $n \ll N$) in the long-range limit $n \rightarrow \infty$. 

Let us consider the quantity,
\be \sigma^z(n) \equiv < s^z_j s^z_{j+n} > \ee
remembering that,
\[ \lim_{n \rightarrow \infty} \sigma^z(n) = <s^z_j>^2 \]
Standard treatise \cite{mattis,LSM,pf} show that for $\phi=\pi$ (no excited
state), this quantity is of the following form. For $\lambda > 1$,
\[ \sigma^z(n) = \left( 1 - \frac{1}{\lambda^2} \right)^{1/4} 
+ A \exp(-\alpha n) \]
($A$, $\alpha$ are constants depending on $\lambda$.) This corresponds to
ferromagnetic state. For $\lambda < 1$, 
\[ \sigma^z(n) =  A^{\prime} \exp(-\alpha^{\prime} n) \] 
($A^{\prime}$, $\alpha^{\prime}$ are constants depending on $\lambda$.) 
This corresponds to paramagnetic state. There is a non-analyticity at the 
point $\lambda = 1$ where the second derivative of ground-state energy 
diverges. 

Let us consider the correlation function 
$C^z(n)$ when $\lambda \neq 1$ and $0 < \phi < \pi$ (i.e. $\phi \neq 0$ and
$\phi \neq \pi$ - some, but not all states are excited). We shall show 
elsewhere \cite{cd2} that this quantity can be calculated exactly using 
Szego's Theorem. The results obtained are as follows.
The quantity 
$\lim_{n \rightarrow \infty} \sigma^z(n)$ will be zero, so that $C^z(n)$ and
$\sigma^z(n)$ are equal. Also, for $\lambda > 1$,
\be C^z(n) = \frac{0.590}{\sqrt{\sin \phi}}
\left( 1-\frac{1}{\lambda^2} \right)^{1/4}
\frac{\cos(\pi+\phi)n}{\sqrt{n}} \ee
and for $\lambda < 1$,
\be C^z(n) = \frac{0.590 \sqrt{\sin \phi}}
{\surd(1+\lambda^2+2\lambda\cos \phi)}
\left( 1-\lambda^2 \right)^{1/4}\frac{1}{\sqrt{n}} \ee

These expressions has been used in the text (Sec. IID, IIIB, IIIC) for the 
analysis of the phase diagram of 1D transverse ANNNI model.


\begin{references}

\bibitem{bkc_book} B.K.~Chakrabarti, A.~Dutta and P.~Sen, {\em Quantum Ising 
Phases and Transitions in Transverse Ising Models} (Springer-Verlag, Berlin, 
Heidelberg) 1996. \\

\bibitem{mattis} D. C. Mattis, {\em The Theory of Magnetism}, Vol. II
(Springer-Verlag, Berlin, Heidelberg) 1985, Sec. 3.6. \\

\bibitem{selke} W. Selke, Phys. Rep. {\bf 170}, 213 (1988). \\ 

\bibitem{rieger} G.~Uimin and H.~Rieger, Z. Phys. B {\bf 101}, 597 (1996). \\

\bibitem{ariz} C.M.~Arizmendi, A.H.~Rizzo, L.N.~Epele and C.A.~Garcia~Canal,
Z. Phys. B {\bf 83}, 273 (1991). \\

\bibitem{sen} P. Sen, S. Chakrabarty, S.~Dasgupta and B.~K.~Chakrabarti,
Z. Phys. B {\bf 88}, 333 (1992). \\

\bibitem{amit}  A.~Dutta and D.~Sen, Phys. Rev. B {\bf 67}, 094435 (2003). \\

\bibitem{selke2} W. Selke and M.E. Fisher, Z. Phys. B {\bf 40}, 71 (1980);
W.~Selke, Z. Phys. B {\bf 43}, 335 (1981). \\

\bibitem{VB} J. Villain and P. Bak, J. Physique {\bf 42}, 657 (1981). \\

\bibitem{jap} T.~Shirahata and T.~Nakamura, Phys. Rev. B {\bf 65}, 024402 
(2001). \\

\bibitem{jap2} R. Derian, A. Gendiar and T. Nishino, cond-mat/0605411. \\

\bibitem{cd1} A. K. Chandra and S. Dasgupta, communicated. \\

\bibitem{LSM} E. Lieb, T. Schultz and D. C. Mattis, Annals of Phys. {\bf 16},
 407 (1961). \\

\bibitem{pf} P. Pfeuty, {\bf 57}, 79 (1970). \\

\bibitem{boundary} The boundary condition for the exact solution is discussed
in Mattis \cite{mattis}, Sec. 3.7. \\

\bibitem{sachdev} S. Sachdev, {\em Quantum Phase Transitions}, (Cambridge 
University Press, Cambridge) 1999, Chapters 1, 3. \\

\bibitem{SDG} S. Dasgupta, Phys. Lett. A {\bf 146}, 181 (1990). \\

\bibitem{cd2} A. K. Chandra and S. Dasgupta, in preparation. \\

\end{references}
\end{document}